\documentclass[letterpaper, 10 pt, conference]{ieeeconf}  

\IEEEoverridecommandlockouts                              
\usepackage{graphics} 
\usepackage{epstopdf} 
\usepackage{epsfig} 
\usepackage{mathptmx} 
\usepackage{times} 
\usepackage{amsmath} 
\usepackage{amssymb}  
\usepackage{dsfont}
\usepackage{color}
\usepackage{todonotes}

\newtheorem{definition}{Definition}

\newtheorem{theorem}{Theorem}
\newtheorem{proposition}{Proposition}
\newtheorem{remark}{Remark}
\newtheorem{lemma}{Lemma}
\newtheorem{corollary}{Corollary}
\newtheorem{assumption}{Assumption}

\title{\LARGE \bf
	Tracking Control  of  Marine Craft in the port-Hamiltonian Framework: A Virtual Differential Passivity Approach}

\author{Rodolfo Reyes-B\'aez, Alejandro Donaire, Arjan van der Schaft, Bayu Jayawardhana, Tristan P\'erez
	\thanks{R. Reyes-B\'aez, B. Jayawardhana and A.J. van der Schaft are with the Jan C. Willems Center for Systems and Control,  Faculty of Science and Engineering, University of Groningen, Nijenborgh 9, 9797AG Groningen, The Netherlands.
		{\tt\small\{r.reyes-baez, b.jayawardhana, a.j.van.der.schaft\}@rug.nl}}%
	\thanks{A. Donaire and T. P\'erez are with the Institute for Future Environments and School of Electrical Eng. and Comp. Sc., Queensland University of Technology, Australia.
		{\tt\small\{alejandro.donaire,         	tristan.perez\}@qut.edu.au}}%
}

\begin{document}

	\maketitle
	\thispagestyle{empty}
	\pagestyle{empty}

	\begin{abstract}
		In this work we propose a family of trajectory tracking controllers for marine craft in the port-Hamiltonian (pH) framework using \emph{virtual differential passivity based control (v-dPBC)}. 
		Two  pH models of marine craft are considered, one in a body frame and another in an inertial frame. The structure and workless forces of  pH models are exploited  to design two virtual control systems  which are related to the original marine craft's pH models. These virtual systems are rendered differentially passive with an imposed steady-state trajectory, both  by means of a control scheme.  Finally,  the original marine craft pH models in closed-loop with above controllers  solve the trajectory tracking problem. The performance of the closed-loop system is evaluated on numerical simulations.
	\end{abstract}

	\section{INTRODUCTION}
	The recent advances of incremental methods in  systems and control have enabled the development  of new control paradigm \cite{slotinecontraction,sontag2010contractive,pavlov2017convergent,jouffroy,forni}. The extension to open systems of differential Lyapunov framework for contraction analysis \cite{forni}, the so-called \emph{differential passivity} \cite{arjan2013differentialpassivity,forni2013differential} resembles the relation between standard passivity property and Lyapunov stability. In other words, the fact that differential passivity implies contraction (with zero input), is analogous to the fact that  passivity implies Lyapunov stability (also with zero input). This gives us the possibility of exploiting systems's interconnection properties in a differential passivity preserving manner \cite{arjan2013differentialpassivity}. Contraction analysis (respectively, differential passivity) has been generalized  by considering contracting (respectively, differentially passive) \emph{virtual systems} \cite{wang,jouffroy,forni} whose input is given by the state of the original systems and whose solutions include all state trajectories of the original system. Roughly speaking, when a virtual system is in cascade with the original system\footnote{In this case, the input of virtual system is connected to the state of original system.} and when both states are initialized at the same point then both systems produce identical state trajectories.  
	
	The latter approach has motivated us to propose \emph{virtual differential passivity-based control} (v-dPBC) methods where we combine the concept of virtual systems and of differential passivity for designing stabilizing/tracking controllers \cite{rodolfochile}. 
	The v-dPBC method consists basically of three main steps. In the first step, we need to define an admissible virtual control system (the precise definition of this will come later). Subsequently, in the second step, we design a controller such that the closed-loop virtual system is differentially passive and has a desired steady-state behaviour or trajectory. Finally, in the third step, we close the loop of the original system using the control law from the previous step where the virtual state is replaced by the original state. The method was applied to fully-actuated \cite{Reyes-BaezIFAC2017} and underactuated  mechanical systems \cite{rodolfochile}. In this paper, we extend these results to marine craft systems which are rigid bodies on \emph{moving frames}.
	
	The dynamic models of marine craft and hydrodynamic forces possess intrinsic passivity properties inherited from its physical nature \cite{fossen2011handbook,fossen1994guidance}. These properties have been widely used for motion control design of ships and underwater vehicles, see for example  \cite{fossen1997nonlinear,SORENSEN1995183,AWOOLSEY20022053}. In the same   spirit, port-Hamiltonian models for marine craft have been proposed  and used for (robust)  passivity-control design  in a structure preserving manner, for further details see \cite{DONAIRE-Tristan,DONAIRE20172167} and reference therein. Specifically, in \cite{DONAIRE20172167}, two pH models of marine vehicle dynamics are presented, one in body-fixed coordinates and another in inertial coordinates. The model in  body-coordinates is later used for designing a robust passivity based tracking control scheme. However, the tracking  problem for the inertial-coordinates model  remains an open problem. From a practical point of view,  the relation of these two models is very useful since the attitude and velocities are usually measured by IMU (Inertial Measurement Unit) and GPS (Global Position System) sensors which means that the  obtained measurements are in the body-fixed frame.
	
	In this work we solve the trajectory  tracking problem of marine craft via the v-dPBC design method for both pH models in  \cite{DONAIRE20172167}. The paper is organized as follows: in Section \ref{sII} we present the  preliminaries that are used through the paper, in Section \ref{sIII}   the notation and standard nomenclature for marine vessels  is introduced together with  a  note on the dynamics in inertial frame that is later used for expressing explicitly the \emph{workless forces} of the marine craft in the body and inertial frame. In Section \ref{sIV} we state the main results on the v-dPBC approach for solving the tracking problem. The performance of a  v-dPBC scheme  is shown  Section \ref{sV}, although due to the limited number of pages, it is briefly described. 

	%
	\section{PRELIMINARIES}\label{sII}
	\subsection{Contraction, differential passivity and virtual systems}
	We adopt \emph{the differential analysis framework} of   \cite{forni}.  Some functions arguments will be omitted due to space limitation.
	\subsubsection{Differential analysis: contraction and passivity}
	Let $\Sigma$ be a nonlinear control system with state space $\mathcal{X}$ be the state-space of dimension $N$, affine in the input $u$, 
	\begin{equation}
	\Sigma_u:\left\{ \begin{array}{llc}
	\dot{{x}}={f}(x,t)+\sum_{i=1}^{n}{g}_i(x,t){u}_i,\\
	y_i=h(x,t), 
	\end{array}
	\right.	
	\label{eq:controlsystem}
	\end{equation}
	where $x\in\mathcal{X}$, ${u}\in\mathcal{U}\subset\mathds{R}^n$ and $y\in\mathcal{Y}$. 
	The vector fields ${f},{g}_i:\mathcal{X}\times\mathds{R}_{\geq 0}\rightarrow T\mathcal{X}$ are  assumed to be smooth and $h:\mathcal{X}\times\mathds{R}_{\geq 0}\rightarrow \mathcal{Y}$. The input space $\mathcal{U}$ and output space $\mathcal{Y}$ are assumed to be open subsets of $\mathds{R}^n$. 	System \eqref{eq:controlsystem} in closed-loop with the smooth static  feedback control law $u= \gamma(x,t)$ will be denoted by 
	\begin{equation}
	\Sigma:\left\{ \begin{array}{llc}
	\dot{{x}}={F}({x},t),\\
	y=h(x,t), 
	\end{array}
	\right.			
	\label{eq:closedloopcontrolsystem}
	\end{equation}
	
	The \emph{variational system} along the  trajectory $(u,x,y)(t)$ is the time-varying system   given by
	\begin{equation}
	\begin{split}
	\left\{ \begin{array}{llc}
	\delta\dot{x}=\frac{\partial f}{\partial x}(x,t)\delta x+\sum_{i=1}^{n}u_i\frac{\partial g_i}{\partial x}(x,t)\delta x+\sum_{i=1}^{n}g_i\delta u_i,\\
	\delta y=\frac{\partial h}{\partial x}(x,t)\delta x.\\
	\end{array}
	\right.		
	\end{split}
	\label{eq:variationalcontrolsystem}
	\end{equation}
	
	\begin{definition}[\cite{crouch}]
		The \emph{prolonged system}  of   $\Sigma_u$ in \eqref{eq:controlsystem}, corresponds to system   described by
		\begin{equation}
		\begin{split}
		\begin{array}{lcc}
		\dot{{x}}={f}({x},t)+\sum_{i=1}^{n}{g}_i({x},t){u}_i,\\
		y=h(x,t),\\
		\delta\dot{x}=\frac{\partial f}{\partial x}(x,t)\delta x+\sum_{i=1}^{n}u_i\frac{\partial g_i}{\partial x}\delta x  +\sum_{i=1}^{n}g_i(x,t)\delta u_i,\\
		\delta y=\frac{\partial h}{\partial x}(x,t)\delta x.\\
		\end{array}
		\end{split}
		\label{eq:prolongedcontrolsystem}
		\end{equation}
		The  prolonged system is \eqref{eq:closedloopcontrolsystem} is
		\begin{equation}
		\begin{split}
		\left\{ \begin{array}{lcc}
		\dot{{x}}={F}({x},t), \\
		{y}={h}(x,t),\\
		\delta\dot{{x}}=\frac{\partial {F}}{\partial{x}}({x},t)\delta{x},\\
		\delta{y}=\frac{\partial {h}}{\partial x}(x,t)\delta x.\\		
		\end{array}
		\right.
		\end{split}
		\label{eq:prolonged-closedloop-system}
		\end{equation}
	\end{definition}		

	\begin{definition}[\cite{forni}]\label{defin:finslerlyapunov}
		A function  $V:T\mathcal{X}\times \mathds{R}_{\geq 0}\rightarrow \mathds{R}_{\geq 0}$ is a candidate \emph{differential or Finsler-Lyapunov function} if it satisfies  the bounds
		\begin{equation}
		c_1 \mathcal{F}({x},\delta{x},t)^{p} \leq V({x},\delta{x},t) \leq c_2\mathcal{F}({x},\delta{x},t)^{p},
		\label{eq:finslerlyapunov}
		\end{equation}
		where $c_1,c_2\in\mathds{R}_{>0}$, $p$ is a positive integer and $\mathcal{F}({x},\cdot,t):=\|\cdot \|_{{x},t}$ is a Finsler structure,  uniformly in $t$.
	\end{definition}

	\begin{definition}\label{defin:finslerdistance}
		Consider a candidate differential Lyapunov function on   $\mathcal{X}$ and the associated Finsler structure $\mathcal{F}$. For any subset $\mathcal{C}\subseteq\mathcal{X}$ and any $x_1,x_2\in\mathcal{C}$, let $\Gamma({x}_1,{x}_{2})$ be the collection of piecewise $C^1$ curves ${\gamma}:I\rightarrow \mathcal{X}$ connecting ${\gamma}(0)={x}_{1}$ and ${\gamma}(1)={x}_{2}$. \emph{The Finsler distance} $d:\mathcal{X}\times\mathcal{X}\rightarrow\mathds{R}_{\geq 0}$ induced by $\mathcal{F}$ is defined by
		\begin{equation}
		d({x}_1,{x}_2):=\inf_{\Gamma({x}_1,{x}_2)}\int_{\gamma}{\mathcal{F}\left({\gamma}(s),\frac{\partial {\gamma}}{\partial s}(s),t\right)}ds.
		\label{defin:finslerdistance1}
		\end{equation}
	\end{definition}
	
	The following result gives a sufficient condition for contraction in terms of  differential Lyapunov functions
	\begin{theorem}\label{theo:lyapunovcontraction}
		Consider  system \eqref{eq:prolonged-closedloop-system}, a connected and forward invariant set $\mathcal{C}\subseteq\mathcal{X}$, and a function $\alpha:\mathds{R}_{\geq 0}\rightarrow \mathds{R}_{\geq 0}$. Let $V$ be a candidate differential Lyapunov function satisfying   
		\begin{equation}
		\dot{V}({x},\delta {x},t)	\leq -\alpha(V({x},\delta {x},t)) 
		\label{eq:finslerlyapunovinequality}
		\end{equation}
		for all $({x},\delta{x})\in T\mathcal{X}$ and all $t>t_0$. Then,  system   \eqref{eq:closedloopcontrolsystem} is 
		\begin{itemize}
			\item  incrementally stable (IS)  if $\alpha(s)=0$ for each $s\geq 0$;
			\item  asymptotically IS if $\alpha$ is of class $\mathcal{K}$;
			\item  exponentially IS with rate $\beta$  if $\alpha(s)=\beta s, \forall s\geq0$.
		\end{itemize}			 
	\end{theorem}
	
	
	\begin{definition}[Contracting system]
		We say that $\Sigma$ contracts  $V$ in $\mathcal{C}$ if \eqref{eq:finslerlyapunovinequality} is satisfied for $\alpha$ of class $\mathcal{K}$. Function $V$ is  the \emph{contraction measure}, and $\mathcal{C}$ is   \emph{the contraction region}.
	\end{definition}

	In analogy to the  standard notion of dissipativity \cite{willem1972,arjanl2}, the differential Lyapunov framework for contraction analysis was extended to open   systems \cite{forni2013differential,arjan2013differentialpassivity}.
	
	\begin{definition}[\cite{arjan2013differentialpassivity}]
		Consider a   control system $\Sigma_u$ in \eqref{eq:controlsystem} together with its prolonged system  \eqref{eq:prolongedcontrolsystem}. Then, $\Sigma_u$  is called \emph{differentially passive} if  there exist a \emph{differential storage function} function ${W}:T\mathcal{X}\rightarrow\mathds{R}_{\geq 0} $ satisfying 
		\begin{equation}
		\frac{d W}{dt}(x,\delta x)\leq \delta y^{\top}\delta u,
		\label{eq:differentialpassivityinequality}
		\end{equation}
		for all $x,\delta x,u,\delta u$. Furthermore,  system \eqref{eq:controlsystem} is called \emph{differentially lossless} if \eqref{eq:differentialpassivityinequality} holds with equality.
	\end{definition}	
	If additionally, the differential storage function is required to be a differential Lyapunov function, then differential passivity implies contraction when the input is $u=0$. 
	\begin{lemma}\label{lemma:dPBC}
		Consider    system $\Sigma_u$ in \eqref{eq:controlsystem}. Suppose there exists a  differential transformation $\delta \tilde{x}=  \Theta(x,t)\delta x$ such that the variational dynamics $\delta\Sigma_u$ in \eqref{eq:variationalcontrolsystem} takes the form
		\begin{equation}
		\begin{split}
		\left\{ \begin{array}{llc}	
		\delta \dot{\tilde{x}}&=\left[\Xi(\tilde{x},t)-\Upsilon(\tilde{x},t)\right]\Pi(\tilde{x},t)\delta\tilde{x}+\Psi(\tilde{x},t)\delta u,\\
		\delta\tilde{y}&=\Psi(t)^{\top}\Pi(\tilde{x},t)\delta \tilde{x},
		\end{array}
		\right.
		\end{split}
		\label{eq:feedbackinterconnectionsystemsvariational}	
		\end{equation}
		where   $\Pi(\tilde{x},t)$ defines a Riemannian metric,   $\Xi(\tilde{x},t)=-\Xi^{\top}(\tilde{x},t)$, $\Upsilon(\tilde{x},t)=\Upsilon^{\top}(\tilde{x},t)$ and $\delta y$ the variational output. If the following inequality holds
		\begin{equation}
		\delta\tilde{x}^{\top}\left[\dot{\Pi}(\tilde{x},t)-2\Pi(\tilde{x},t)\Upsilon(\tilde{x},t)\Pi(\tilde{x},t)\right]\delta\tilde{x}\leq-\alpha(W)
		\label{eq:dPBCinequality}
		\end{equation}
		with  $\alpha$   of class $\mathcal{K}$. Then, $\Sigma_u$ is differentially passive from   $\delta u$ to $\delta\tilde{y}$ with respect to the differential storage function 
		\begin{equation}
		W(\tilde{x},\delta \tilde{x})=\frac{1}{2}\delta \tilde{x}^{\top}{\Pi}(\tilde{x},t)\delta \tilde{x}.
		\label{eq:differentialStorage-Lemma}
		\end{equation}	
	\end{lemma}
	
	\subsubsection{Contracting  virtual systems} 
	A generalization of contraction   was  introduced in \cite{wang}   to study  the  convergence between  solutions of two or more    systems. This concept is based on the contraction behavior of a   \emph{virtual system}.
	
	\begin{definition}[Virtual system]\label{defin:virtualsystem1}
		Consider systems $\Sigma$ and $\Sigma_u$, given by \eqref{eq:closedloopcontrolsystem} and \eqref{eq:controlsystem}, respectively. Suppose that  $\mathcal{C}_v\subseteq \mathcal{X}$ and $\mathcal{C}_x\subseteq \mathcal{X}$ are connected and forward invariant sets. 
		A \emph{virtual system} associated to $\Sigma$ is defined as the system 
		\begin{equation}
		{\Sigma}_{v}:\left\{ \begin{array}{lcc}
		\dot{{x}}_v= {\Phi}_v(x_v,{x},t), \\
		{y}_v={h}_v(x_v,x,t),		
		\end{array}
		\right.		
		\label{eq:virtualsystemTheorem}
		\end{equation}
		in the state $x_v\in\mathcal{C}_v$ and  parametrized by $x\in\mathcal{C}_x$, where   $\Phi:\mathcal{C}_v\times\mathcal{C}_x\times\mathds{R}_{\geq 0}\rightarrow T\mathcal{X}$ and $h_v:\mathcal{C}_v\times\mathcal{C}_x\times\mathds{R}_{\geq 0}\rightarrow T\mathcal{X}$  satisfy  the condition
		\begin{equation}
		{\Phi}_v({x},{x},t)=F(x,t)\quad \text{and}\quad h_v(x,x,t)=h(x,t),  
		\label{eq:virtualcontrolsystemTheorem}		
		\end{equation}
		for all $t\geq t_0$. Similarly, a  \emph{virtual control system}  for $\Sigma_u$ is defined as the control system 
		\begin{equation}
		\begin{split}
		{\Sigma}_{uv}:\left\{ \begin{array}{lcc}
		\dot{x}_v=\Gamma(x_v,x,u,t),\\
		{y}_v={h}_v(x_v,x,t), \quad\quad  \forall t\geq t_0,		
		\end{array}
		\right.			
		\end{split}
		\label{eq:virtualsystemTheorem1}	
		\end{equation}
		in  the state $x_v\in\mathcal{X}$ and parametrized by   $x\in\mathcal{X}$, the output $y_v\in\mathcal{Y}$, where  $h_v:\mathcal{C}_v\times\mathcal{C}_x\times\mathds{R}_{\geq 0}\rightarrow  \mathcal{Y}$ and  $\Gamma:\mathcal{C}_v\times\mathcal{C}_x\times\mathds{R}_{\geq 0}\rightarrow T \mathcal{X}$ satisfy 
		\begin{equation}
		\begin{split}
		\Gamma(x,x,u,t)&=f(x,t)+G(x,t)u,\\
		h_v(x,x,t)&=h(x,t),\quad\quad\quad \forall u, \forall t\geq t_0.		
		\end{split}
		\end{equation}		
	\end{definition}	
	
	\begin{theorem}[Virtual contraction \cite{wang,forni2013differential}]\label{theo:partialcontraction}
		Consider  $\Sigma$ and $\Sigma_v$ in \eqref{eq:closedloopcontrolsystem} and \eqref{eq:virtualsystemTheorem}, respectively. Let  $\mathcal{C}_v\subseteq \mathcal{X}$ and $\mathcal{C}_x\subseteq \mathcal{X}$ be two connected and forward invariant sets. Suppose that $\Sigma_v$  is uniformly contracting with respect to $x_v$. Then, for any  $x_0\in\mathcal{C}_x$ and $x_{v0}\in\mathcal{C}_v$, each solution to $\Sigma_v$ asymptotically converges   to the solution  of $\Sigma$.
	\end{theorem}	
	If Theorem \ref{theo:partialcontraction} holds, then the actual system $\Sigma$ is said to be \emph{virtually contracting}\footnote{The concept was originally called \emph{partial contraction} in \cite{wang}. However,  to avoid confusion with future research we use the adjective virtual.}.  In case of the virtual control system $\Sigma_{uv}$ is differentially passive,  then  the  actual control system $\Sigma_u$ is said to be \emph{virtually differentially passive}.

	\subsubsection{Virtual differential passivity based control (v-dPBC)}\label{subsection:v-dPBC}				
	The  design method\footnote{Virtual systems for control design were already considered in \cite{jouffroy,manchester2015unifying}.}  is divided in three main steps:
	\begin{itemize}
		\item Design the virtual   system control \eqref{eq:virtualsystemTheorem1} for system  \eqref{eq:controlsystem}.
		\item Design the feedback $u=\zeta(x_v,x,t)+\omega$ for \eqref{eq:virtualsystemTheorem1} such that the closed-loop virtual system   is differentially  passive for  the input-output pair  $(\delta {{y}}_v,\delta\omega)$ and has a desired trajectory $x_d(t)$ as steady-state   solution.
		\item Define the controller for  system \eqref{eq:controlsystem} as $u=\eta(x,x,t)$.
	\end{itemize}
	
	%

	\subsubsection{Trajectory tracking control via v-dPBC}\label{subsection:trackingproblem}	
	Above method  can be directly applied to solve the trajectory tracking problem, which for system \eqref{eq:controlsystem} is   stated as follows:		
	
	\emph{Tracking   problem:} Given   $x_d(t)$,  design a control law $u(x,t)$ for   system \eqref{eq:controlsystem} such that $x(t)\rightarrow x_d(t)$ as $t\rightarrow \infty$, uniformly.
	Solution: in the second step of v-dPBC  split the control   as  
	\begin{equation}
	\eta(x_v,x,t):=u_{ff}(x_v,x,t)+u_{fb}(x_v,x,t)
	\label{eq:control-tracking}
	\end{equation}
	such that
	\begin{itemize}
		\item The \emph{feedforward-like} term $u_{ff}$ ensures that the closed-loop virtual  system has    the desired  trajectory $x_d(t)$ as particular solution.
		\item The \emph{feedback} action $u_{fb}$ commands the closed-loop system  to be differentially passive in a connected and forward complete set $\mathcal{C}\subseteq\mathcal{X}$. 		
	\end{itemize}	
	
	\subsection{Mechanical port-Hamiltonian and virtual systems}	
	
	Previous ideas will be applied to mechanical pH systems.
	\begin{definition}[\cite{vanderschaft1995}]\label{def:pHsystem}
		A {port-Hamiltonian system} with  $N$ dimensional state space  $\mathcal{X}$, input and output spaces $\mathcal{U}=\mathcal{Y}\subset\mathds{R}^{m}$ and Hamiltonian function $H:\mathcal{X}\rightarrow\mathds{R}$, is given by
		\begin{equation}
		\begin{split}
		\dot{x}&=\left[J_H(x)-R(x)\right]\frac{\partial H}{\partial x}(x)+G(x)u\\			
		y&= G^{\top}(x)\frac{\partial H}{\partial x}(x),
		\end{split}
		\label{eq:IOpHsystem}			
		\end{equation}
		where $G(x)$ is a $N\times m$ input matrix, and   $J_H(x)$, $R(x)$ are the interconnection and dissipation $N\times N$ matrices which satisfy $J_H(x)=-J_H^{\top}(x)$ and $R(x)=R^{\top}(x)\geq0$. 
	\end{definition}
	
	In the specific case of a standard mechanical system with generalized coordinates $q$ on the configuration space $\mathcal{Q}$ of dimension $n$ and  velocity $\dot{{q}}\in T_{q}\mathcal{Q}$, the Hamiltonian function is given by the total energy
	\begin{equation}
	H({x})=\frac{1}{2}{p}^{\top}{M}^{-1}({q}){p}+P({q}),
	\label{eq:phmechanicalHamiltonian}
	\end{equation}
	where ${x}=({q},{p})\in T^*\mathcal{Q}:=\mathcal{X}$ is the   state, $P({q})$ is the potential energy,  $p:=M(q)\dot{q}$ is the   momentum and the inertia matrix $M(q)$ is symmetric and positive definitive. Then,  the pH system \eqref{eq:IOpHsystem} takes the form
	\begin{equation}
	\begin{split}
	\begin{bmatrix}
	\dot{q}\\
	\dot{p}
	\end{bmatrix}&=\begin{bmatrix}
	0 & I\\
	-I & -D(q)
	\end{bmatrix}\begin{bmatrix}
	\frac{\partial H}{\partial q}(q,p)\\
	\frac{\partial H}{\partial p}(q,p)
	\end{bmatrix}+\begin{bmatrix}
	0\\
	B(q)
	\end{bmatrix}u,\\
	y&=B^{\top}(q)\frac{\partial H}{\partial p}(q,p),
	\end{split}
	\label{eq:Hamiltonian2}
	\end{equation}
	with matrices
	\begin{equation}
	J_H(x)=\begin{bmatrix}
	{0} & {I}\\
	-{I}& {0}
	\end{bmatrix};  R(x)=\begin{bmatrix}
	{0} & {0}\\
	{0}& {D}({{q}})
	\end{bmatrix}; G(x)=\begin{bmatrix}
	{0}\\
	B
	\end{bmatrix},
	\label{eq:phmechanical}		
	\end{equation}		
	where  ${D}({q})={D}^{\top}({q}) \geq {0}$ is the damping matrix and $I$ and $0$ are the $n\times n$ identity, respectively, zero matrices. The input force matrix $B(q)$ has rank $m\leq n$.
	\section{Marine Craft's port-Hamiltonian modeling}\label{sIII}
	In this part we adopt the modeling framework in reference \cite{fossen2011handbook} and the notation of \emph{SNAME (1950)} for marine vessels. From a guidance, navigation and control point of view, in the modeling of marine craft, four different coordinate frames are considered: the Earth Centered inertial (ECI) frame $\{i\}$, whose origin $O_i$ is located at the center of mass of the Earth; the Earth-centered-Earth-fixed (ECEF) frame $\{e\}$ that rotates with the Earth;   the North-East-Down (NED) frame $\{n\}$ with origin $O_n$ defined relative to the Earth's reference ellipsoid (WGS84); and the body frame $\{b\}$ which is a moving coordinate frame that is fixed to the craft\footnote{For a marine craft, the origin $O_b$ is usually chosen to coincide with a point midship $CO$ in the water line; while  the body-frame axes are chosen to coincide with the principal axes of inertia \cite{fossen2011handbook}.}.  In this work we take the following modeling  assumptions:
	\begin{assumption}[Flat navigation]
		The operating radius of a marine craft is limited.  We assume $\{n\}$ to be inertial.
	\end{assumption}
	\begin{assumption}[Maneuvering theory]
		The hydrodynamic coefficients are frequency independent (no wave excitation). 	
	\end{assumption}
	
	Under above assumptions, the equation of motion in body-fixed coordinates are \cite{fossen2011handbook}: 
	\begin{equation}
	\begin{split}
	\dot{\eta}&=J(\eta)\nu,\\
	M\dot{\nu}+C(\nu)\nu+D(\nu)\nu+g(\eta)&=\tau
	\end{split}
	\label{eq:FossenmodelLagrange}
	\end{equation}
	where $\eta=[x,y,z,\phi,\theta,\psi]^{\top}\in\mathcal{Q}:=\mathds{R}^{3}\times S^{3}$ describes the craft's position and orientation; $\nu=[\nu_1^{\top},\nu_2^{\top}]^{\top}$ is the  (quasi-)velocity in $\{b\}$, with $\nu_1[u,v,w]^{\top}$(surge, sway, heave)  and $\nu_2=[p,q,r]^{\top}$(roll, pith, yaw); and $\tau$ are the force and torque inputs. The    matrix function $J(\eta)$ is a well defined    transformation if $\theta\neq \pm \pi/2$, due to the Euler angles representation  \cite{fossen2011handbook}.  The inertia matrix $M=M^{\top}>0$ is given by
	\begin{equation}
	M:=\begin{bmatrix}
	m I & -mS_{\times}(r_g^b)\\
	mS_{\times}(r_g^b) & I_b
	\end{bmatrix}
	\end{equation}
	where $m$ is the total mass due to  the craft's mass and fluid added mass, $I_b$ the moment of inertia in $\{b\}$, $r^{b}_g$  the constant vector between $O_b$ and the center of gravity (CG) in $\{b\}$-coordinates, and the matrix $S_{\times}(\cdot)$ is defined as 
	\begin{equation}
	S_{\times}(a)=\begin{bmatrix}
	0 & -a_3 & a_2\\
	a_3 & 0 & -a_1\\
	-a_2 & a_1 & 0
	\end{bmatrix}.
	\end{equation}
	According to Kirchhoff's equations of motion, the Coriolis-centripetal matrix $C(\nu)$ is any matrix satisfying 
	\begin{equation}
	C(\nu)\nu=\begin{bmatrix}
	S_{\times}(\nu_2)\frac{\partial T}{\partial \nu_1}(\nu)\\
	S_{\times}(\nu_2)\frac{\partial T}{\partial \nu_2}(\nu)+S_{\times}(\nu_1)\frac{\partial T}{\partial \nu_1}(\nu)
	\end{bmatrix},
	\label{eq:Coriolis-centripetal-Body}
	\end{equation}
	where $T^*(\nu)=\frac{1}{2}\nu^{\top}M\nu$ is the kinetic (co-)energy; $D(\nu)=D^{\top}(\nu)>0$ is the total hydrodynamic damping matrix; and $g(\eta)$ is the vector of hydrostatic forces and torques due to gravity and buoyancy.	
	
	The force   $F^b_{gr}(\nu)=C(\nu)\nu$ in \eqref{eq:Coriolis-centripetal-Body} is \emph{workless} for every $\nu$. 
	Following \cite{chang2002controlled}, there exists a skew-symmetric matrix $S_{L}^b(\nu)$ such that $F_{gr}^b(\nu)=S_{L}^b(\nu)\nu$. This was also showed in  \cite{fossen2011handbook} for the specific  case of system \eqref{eq:FossenmodelLagrange}.

	\subsection{A note on marine craft's dynamics in the inertial frame}
	The equations of motion \eqref{eq:FossenmodelLagrange} in the body-frame $\{b\}$ can be expressed in the inertial frame $\{n\}$ by performing the kinematic transformation $\dot{\eta}=J(\eta)\nu$ as follows \cite[p.48]{fossen1994guidance}:
	\begin{equation}
	\begin{split}
	M_{\eta}(\eta)\ddot{\eta}+C_{\eta}(\eta,\dot{\eta})\dot{\eta}+D_{\eta}(\eta,\dot{\eta})\dot{\eta}+g_{\eta}(\eta)=\tau_{\eta}
	\end{split}
	\label{eq:FossenmodelLagrange-intertial}
	\end{equation}
	where
	\begin{equation}
	\begin{split}
	M_{\eta}(\eta)&=J^{-\top}(\eta)MJ^{-1}(\eta),\\
	C_{\eta}(\eta,\dot{\eta})&=J^{-\top}\left[C(J^{-1}\dot{\eta})-MJ^{-1}(\eta)\dot{J}(\eta)\right]J^{-1},\\
	D_{\eta}(\eta,\dot{\eta})&=J^{-\top}(\eta)D(J^{-1}\dot{\eta})J^{-1}(\eta),\\
	g_{\eta}(\eta)&=J^{-\top}(\eta)g(\eta),\\
	\tau_{\eta}&=J^{-\top}(\eta)\tau.
	\end{split}
	\label{eq:matricesequations-inertial}
	\end{equation}
	Alternatively, the Lagrange equations can be used to derive the model \eqref{eq:FossenmodelLagrange-intertial} since in $\{n\}$ the generalized position vector $\eta$ and its corresponding  generalized velocity $\dot{\eta}$ can be used to express the kinetic (co-)energy as
	\begin{equation}
	T_{\eta}^{*}(\eta,\dot{\eta})=\frac{1}{2}\dot{\eta}^{\top}M_{\eta}(\eta)\dot{\eta},
	\end{equation}
	and hence the Lagrangian function $L(\eta,\dot{\eta})=T^{*}_{\eta}-P(\eta)$ is well defined. For the dissipative hydrodynamic forces a Rayleigh  function $F_{R}(\eta,\dot{\eta})=\frac{1}{2}\dot{\eta}^{\top}D_{\eta}(\eta,\dot{\eta})\dot{\eta}$ is considered. With the Lagrangian approach, the Coriolis-centripetal matrix   is any matrix $C_{\eta}(\eta,\dot{\eta})$ satisfying
	\begin{equation}
	C_{\eta}(\eta,\dot{\eta})\dot{\eta}=\dot{M}_{\eta}(\eta)\dot{\eta}-\frac{\partial}{\partial \eta}\left(\frac{1}{2}\dot{\eta}^{\top}M_{\eta}(\eta)\dot{\eta}\right).
	\label{eq:CoriolisInertial}
	\end{equation}
	which resembles the definition of the Coriolis-centripetal matrix in the standard Euler-Lagrange setting in \cite{ortega2013passivity}.  From \eqref{eq:CoriolisInertial} is is easy to see that 
	\begin{equation}
	\dot{\eta}^{\top}\underbrace{\left[\frac{1}{2}\dot{M}_{\eta}(\eta)\dot{\eta}-\frac{\partial}{\partial \eta}\left(\frac{1}{2}\dot{\eta}^{\top}M_{\eta}(\eta)\dot{\eta}\right)\right]}_{F_{gr}(\eta,\dot{\eta})} =0.
	\end{equation}
	Thus,  the force $F_{gr}(\eta,\dot{\eta})$ is workless which implies that there exists a skew-symmetric matrix $S^{L}_{\eta}(\eta,\dot{\eta})$ such that $F_{gr}(\eta,\dot{\eta})=S^{L}_{\eta}(\eta,\dot{\eta})\dot{\eta}$. Thus,  \eqref{eq:CoriolisInertial} takes the form
	\begin{equation}
	C_{\eta}(\eta,\dot{\eta})\dot{\eta}=\left[\frac{1}{2}\dot{M}_{\eta}(\eta)+S^{L}_{\eta}(\eta,\dot{\eta})\right]\dot{\eta}.
	\label{eq:Cortiolis-Smatrix}
	\end{equation}
	With this form, it is straightforward to verify that $$\dot{\eta}^{\top}\left[\dot{M}_{\eta}(\eta)-2C_{\eta}(\eta,\dot{\eta})\right]\dot{\eta}=0,$$
	for all $\dot{\eta}$ and any $C_{\eta}(\eta,\dot{\eta})$. In case the matrix $C_{\eta}(\eta,\dot{\eta})$ is expressed in terms of the  Christoffel symbols of $M_\eta(\eta)$, the skew-symmetric matrix $S^{L}_{\eta}(\eta,\dot{\eta})$ is given as 
	\begin{equation}
	\begin{split}
	S_{\eta kj}^L(q,\dot{q})
	&= \frac{1}{2} \sum_{i=1}^{n}\left\{\frac{\partial M_{\eta ki}}{\partial q_j}(q)  -\frac{\partial  M_{\eta ij}}{\partial q_k}(q) \right\}\dot{q}_i.
	\label{eq:SlLagrangian}
	\end{split}
	\end{equation}
	From above observations and $C_{\eta}(\eta,\dot{\eta})$ in \eqref{eq:matricesequations-inertial}, motivates another   choice of $S^{L}_{\eta}(\eta,\dot{\eta})$ as
	$$S^{L}_{\eta}=J^{-\top} C(J^{-1}\dot{\eta})J^{-1} +\frac{1}{2}\left[{J}^{-\top}M\dot{J}^{-1}-({J}^{-\top}M\dot{J}^{-1})^{\top}\right].$$
	\begin{remark}
		In   \cite[p.54]{fossen1994guidance}, there is not a clear relation between the Coriolis-centripetal matrix  obtained with the Euler-Lagrange equations   and the one in \eqref{eq:matricesequations-inertial}. However, due to identity \eqref{eq:Cortiolis-Smatrix},   this relation is made clear.
	\end{remark}		
	
	\subsection{Craft's pH model and  virtual system in body-frame}
	In order to get a pH model of  \eqref{eq:FossenmodelLagrange}, in  \cite{DONAIRE-Tristan} and more recently in  \cite{DONAIRE20172167}, the next assumption is made:
	\begin{assumption}\label{assumptionDonaire}
		There exists  $P:\mathcal{Q}\rightarrow \mathds{R}$  satisfying 
		\begin{equation}
		J^{\top}(\eta)\frac{\partial P}{\partial \eta}(\eta)=G(\eta).
		\end{equation}
	\end{assumption}
	Under Assumption \ref{assumptionDonaire}, the marine craft  dynamics  \eqref{eq:FossenmodelLagrange} can be written in  port-Hamiltonian form as follows \cite{DONAIRE20172167}: 
	\begin{equation}
	\begin{bmatrix}
	\dot{\eta}\\
	\dot{p}^b 
	\end{bmatrix}=\begin{bmatrix}
	0 & J(\eta)\\
	-J^{\top}(\eta) & -J_2({p}^b )
	\end{bmatrix}\begin{bmatrix}
	\frac{\partial H}{\partial \eta}(\eta,{p}^b )\\
	\frac{\partial H}{\partial \dot{p}^b }(\eta,{p}^b )
	\end{bmatrix}+\begin{bmatrix}
	0\\
	I_n
	\end{bmatrix}\tau,
	\label{eq:pH-body}
	\end{equation}
	where the  Hamiltonian function is
	\begin{equation}
	H(\eta,\dot{p}^b )=\frac{1}{2}{p}^{b\top}M^{-1}{p}^b +P(\eta), 
	\label{eq:pH-body-Hamiltonian}
	\end{equation}
	the \emph{quasi momentum} is defined as\footnote{Since $p^b$ is not the true momentum of  $\eta$ and $\nu$ are  velocities in $\{b\}$ (\emph{quasi-velocities}), for details see \cite[p. 193]{greenwood} and \cite{fossen2011handbook}.}
	\begin{equation}
	{p}^b =M\nu,
	\label{eq:momentum-body-frame}
	\end{equation}
	and $J_2({p}^b )=C(M^{-1}{p}^b )+D(M^{-1}{p}^b )$. It is easy to see that \eqref{eq:pH-body} is  passive with \eqref{eq:pH-body-Hamiltonian} as storage function.   Define $E^b(\eta,{p}^b ): =C(M^{-1}{p}^b )$ for future purposes. 
	\subsection{Craft's pH model and  virtual system in inertial-frame}
	Also in \cite{DONAIRE20172167}, under Assumption \ref{assumptionDonaire}, and  pH model in inertial coordinates was developed by translating the (quasi)momentum vector $p^b$ in $\{b\}$ to the frame  $\{n\}$ as  
	\begin{equation}
	p=J^{-\top}(\eta)p^b {\iff p=M_{\eta}(\eta,\dot{\eta})\dot{\eta}}.
	\label{eq:Diffeo-body-inertial}
	\end{equation}
	Then, the dynamics \eqref{eq:momentum-body-frame} in coordinates $(\eta,p)$ is \cite{DONAIRE20172167}:
	\begin{equation}
	\begin{split}
	\begin{bmatrix}
	\dot{\eta}\\
	\dot{p}
	\end{bmatrix}
	&= \begin{bmatrix}
	0 & I\\
	-I & -L(\eta,p)
	\end{bmatrix}\begin{bmatrix}
	\frac{\partial H_{\eta}}{\partial \eta}(\eta,p)\\
	\frac{\partial H_{\eta}}{\partial p}(\eta,p)\\
	\end{bmatrix}+\begin{bmatrix}
	0\\
	J^{-\top}
	\end{bmatrix}\tau,
	\end{split}
	\label{eq:pH-inertial}
	\end{equation}
	where the Hamiltonian function is  
	\begin{equation}
	H_{\eta}(\eta,p)=\frac{1}{2}p^{\top}M_{\eta}^{-1}(\eta)p+P(\eta),
	\label{eq:pH-inertial-Hamiltonian}
	\end{equation}
	and the matrix
	\begin{equation}
	L(\eta,p)=\left[\frac{\partial J^{-\top}}{\partial \eta}J^{\top}p\right]^{\top}-\frac{\partial J^{-\top}}{\partial \eta}J^{\top}p+J^{-\top} J_2 (M^{-1}J^{\top}p)J^{-1}.
	\end{equation}
	Notice that \eqref{eq:pH-inertial} is a port-Hamiltonian system provided that it satisfies Definition \ref{def:pHsystem}. However, this system is not  a standard mechanical pH system like \eqref{eq:Hamiltonian2}, since its  interconnection matrix  can not be written as the canonical one in \eqref{eq:phmechanical}. Nevertheless, system \eqref{eq:pH-inertial}  is still mechanical since its interconnection matrix defines an \emph{almost-Poisson structure}\footnote{The associated Poisson bracket does not satisfy the Jacobi identity \cite{chang2002controlled}.} \cite{chang2002controlled}, that is, the energy conservation   is satisfied. In the following proposition an alternative form of   \eqref{eq:pH-inertial} is presented.
	
	\begin{proposition}\label{prop:alternative-form-inertial}
		The pH system  \eqref{eq:pH-inertial} takes the   form
		\begin{equation}
		\begin{split}
		\begin{bmatrix}
		\dot{\eta}\\
		\dot{p}
		\end{bmatrix}
		&= \begin{bmatrix}
		0 & I\\
		-I & -(E_{\eta}(\eta,p)+D_{\eta}^H(\eta,p))
		\end{bmatrix}\begin{bmatrix}
		\frac{\partial P}{\partial \eta}(\eta)\\
		M^{-1}_{\eta}p
		\end{bmatrix}+\begin{bmatrix}
		0\\
		J^{-\top}
		\end{bmatrix}\tau,
		\end{split}
		\label{eq:pH-inertial-alternative}
		\end{equation}
		where  
		\begin{equation}
		\begin{split}
		E&=S^{H}_{\eta}(\eta,p)-\frac{1}{2}\dot{M}_{\eta}(\eta),\\
		S^{H}_{\eta}&=J^{-\top}CJ^{-1}+\frac{1}{2}\left[\left(\frac{\partial J^{-\top}}{\partial \eta}J^{\top}p\right)^{\top}-\frac{\partial J^{-\top}}{\partial \eta}J^{\top}p\right],\\
		D_{\eta}^H&=D_{\eta}(\eta,M^{-1}_{\eta}(\eta)p).
		\end{split}
		\end{equation}	
	\end{proposition}

	\section{Trajectory Tracking Control Design}\label{sIV}
	
	\subsection{Control design in the body-fixed frame}
	\subsubsection{Virtual control system design}	
	The structure of  \eqref{eq:pH-body} motivates the definition of a virtual system for it, in the state $(\eta_v,p^b_v)$ and parametrized by the trajectory $(\eta,p^b)$, as
	\begin{equation}
	\begin{split}
	\begin{bmatrix}
	\dot{\eta}_v\\
	\dot{p}^b _v
	\end{bmatrix}=\begin{bmatrix}
	0 & J(\eta)\\
	-J^{\top}(\eta) & -J_2(p^b)
	\end{bmatrix}\begin{bmatrix}
	\frac{\partial H_v}{\partial \eta_v}(\eta_v,p^b_v)\\
	\frac{\partial H_v}{\partial p^b_v}(\eta_v,p^b_v)
	\end{bmatrix}+\begin{bmatrix}
	0\\
	I_n
	\end{bmatrix}\tau,
	\end{split}
	\label{eq:pH-body-virtual}
	\end{equation}
	with  
	\begin{equation}
	H_v(\eta_v,p^b_v)=\frac{1}{2}p_v^{b\top}M^{-1}p^b_v+P_{v}(\eta_v)
	\label{eq:pH-body-virtual-Hamiltonian}
	\end{equation}
	where $P_{v}(\eta_v)$ also fulfills Assumption \ref{assumptionDonaire} and $P_v(\eta)=P(\eta)$. 
	Remarkably,   \eqref{eq:pH-body-virtual} is also a  pH system and   passive  with \eqref{eq:pH-body-virtual-Hamiltonian} as storage function and supply rate $y_{b\eta}^{\top}\tau_{\eta}$, where   $y^b_{\eta}=M^{-1}p_v^b$, uniformly in  $(\eta(t),p^b(t)$, for all $t>t_0$.
	
	\subsubsection{Differential passivity based control design}
	In this step of the method v-dPBC, we will design  $\tau=\zeta^b(x^b,x^b_v,t)+\omega^b$ such that \eqref{eq:pH-body-virtual} is differentially passive in the closed-loop.
	\begin{proposition}\label{proposition:fullstatecontroller-body}
		Consider system \eqref{eq:pH-body-virtual} and a smooth  trajectory ${x}_d^b=(\eta_d,p_d^b)$ in   $\{b\}$ with $p^b_d=M^{-1}\dot{\eta}_d$. Let us introduce the following error coordinates 
		\begin{equation}
		\tilde{{x}}_v^b:=\begin{bmatrix}
		\tilde{{\eta}}_v^b\\
		{\sigma_v^b}
		\end{bmatrix}=\begin{bmatrix}
		{\eta}_v-{\eta}^b_d \hfill \\
		{p}_v^b-{p}_{r}^b 
		\end{bmatrix},
		\label{eq:changeofcoordinatesError}
		\end{equation}
		where the auxiliary momentum reference ${p}_{r}^b $ is given by
		\begin{equation}
		p^b_{r}(\tilde{\eta}_v^b,t):=M(\dot{\eta}^b_d-\phi^b(\tilde{\eta}^b_v)),
		\label{eq:auxiliarreference}
		\end{equation}
		$\phi^b:\mathcal{Q}\rightarrow T\mathcal{Q}$ is such that $\phi^b(0)=0$; and $\Pi_{{\eta}_v}^b:\mathcal{Q}\times \mathds{R}_{\geq 0}\rightarrow \mathds{R}^{n\times n}$ is a positive definite Riemannian metric tensor  satisfying the inequality
		\begin{equation}
		\begin{split}
		\dot{\Pi}_{{\eta}_v}^b&(\tilde{\eta}^b_v,t)-\Pi_{{\eta}_v}^b(\tilde{\eta}^b_v,t)\frac{\partial \phi^b}{\partial \tilde{\eta}_v^b}(\tilde{\eta}_v^b)\\&-\frac{\partial \phi^{b\top}}{\partial \tilde{\eta}_v^b}(\tilde{\eta}_v^b)\Pi_{{\eta}_v}^b(\tilde{\eta}^b_v,t)\leq  -2\beta_{\eta_v}^b(\tilde{\eta}_v^b,t)\Pi_{{\eta}_v}^b(\tilde{\eta}^b_v,t),
		\end{split}
		\label{eq:controllaw-positionmetricinquality}
		\end{equation}							
		with $\beta_{\eta_v}^b(\tilde{\eta}_v^b,t)>0$, uniformly. Assume also that the $i$-th row of  $\Pi_{{\eta}_v}^b(\tilde{\eta}^b_v,t)$  is a conservative vector field\footnote{This   ensures that the integral  in  \eqref{eq:controlaw1} is well defined and independent of the path connecting  $0$ and  $\tilde{q}_v$.}. Consider  also the composite control law given by
		\begin{equation}
		{\tau}(x_v^b,x^b,t):={\tau}_{ff}(x_v^b,x^b,t)+{\tau}_{fb}(x_v^b,x^b,t)+ \omega^b,
		\label{eq:controlaw}
		\end{equation}
		where $x^b=(\eta,p^b)$, $x_v^b=(\eta_v,p^b_v)$ and
		\begin{equation}
		\begin{split}
		{\tau}_{ff}&=\dot{p}^b_{r}+\frac{\partial P_v}{\partial \eta_v}+\big[E^b(\eta,p^b)+D(M^{-1}p^b)\big] M^{-1}p_{r}^b,\\
		{\tau}_{fb}&=-\int_{0}^{\tilde{{\eta}}_v^b}\Pi_{{\eta}_v}^b(\xi,t)d\xi-{K}_dM^{-1}\sigma_v^b,
		\end{split}
		\label{eq:controlaw1}
		\end{equation}
		with ${K}_d>0$ and $\omega^b$ be an external input. Then,
		\begin{enumerate}
			\item   system \eqref{eq:pH-body-virtual} in closed-loop with \eqref{eq:controlaw}  is  differentially passive from  $\delta \omega^b$ to $\delta \overline{y}_{\sigma_v}^b=M^{-1}\delta\sigma_v^b$ with respect to the differential storage function
			\begin{equation}
			W_{x^b}(\tilde{x}_v^b,\delta \tilde{x}_v^b)=\frac{1}{2}\delta \tilde{x}_v^{b\top}\begin{bmatrix}
			\Pi_{{\eta}_v^b}& 0\\
			0 & M^{-1}
			\end{bmatrix}\delta \tilde{x}_v^b.
			\label{eq:design-dLCF-sigma}
			\end{equation}			
			\item the closed-loop variational dynamics   preserves the  structure of \eqref{eq:feedbackinterconnectionsystemsvariational}, with 
			\begin{equation}
			\begin{split}
			\Pi^b&=\text{diag}\{\Pi_{{\eta}_v}^b(\tilde{\eta}^b_v,t),M^{-1}\}\\
			\Upsilon^b&=\text{diag}\bigg\{\frac{\partial \phi^b}{\partial \tilde{\eta}_v^b}\Pi_{{\eta}_v}^{b-1},D(M^{-1}p^b)+K_d \bigg\},\\
			\Xi^b&=\begin{bmatrix}
			0  & I\\
			-I & -S_H^b(p^b)\\		
			\end{bmatrix}; \quad \Psi^b=\begin{bmatrix}
			0\\I
			\end{bmatrix}.		
			\end{split}
			\label{eq:feedbackinterconnectionsystemsvariational-flexible}
			\end{equation}
		\end{enumerate}
	\end{proposition}

	\begin{remark}
		The \emph{timed-IDA-PBC} method  proposed  in  \cite{iran} uses the contraction property of a target pH system for solving the tracking problem. In particular, one of the main assumptions is that the interconnection and dissipation matrices are constant. In this regards, our proposed v-dPBC approach can be used to relax this assumption for the marine craft case with the  pH system \eqref{eq:pH-body}.
	\end{remark}
	
	\subsubsection{Actual system's controller}
	Here we show that the controller which is designed in the previous item can be used as a trajectory tracking controller for the actual  Hamiltonian system \eqref{eq:pH-body}. This is stated in the following corollary.
	\begin{corollary}\label{corollary:Actualtrackingcontrol}
		Consider the controller \eqref{eq:controlaw}. Then, all solutions of system \eqref{eq:pH-body}  in closed-loop with the control law  $$\tau(x^b,x^b,t)=\tau_{ff}(x^b,x^b,t)+\tau_{fb}(x^b,x^b,t)$$
		converge exponentially to the  trajectory $x^b_d(t)$  with rate 
		\begin{equation}
		\begin{split}
		\beta^b&=\min\{\beta_{\eta_v}^b,\lambda_{\min}\{D+K_d\}\lambda_{\min}\{M^{-1}\}\},
		\label{eq:trackingrate} 
		\end{split}
		\end{equation}
		where $\lambda_{min}(\cdot)$  is the minimum  eigenvalue of  its argument.
	\end{corollary}
	\subsection{Control design in the inertial frame}
	The control design procedure in the inertial frame is  almost identical to that in the body-fixed frame. Thus, due to space limitation we only state the results without proof.
	\subsubsection{Virtual control system design} 
	With the alternative form \eqref{eq:pH-inertial-alternative} in Proposition \ref{prop:alternative-form-inertial}, we define a virtual system  associated with the pH system  \eqref{eq:pH-inertial}, in the state $(\eta_v,p_v)$ and parametrized by $(\eta,p)$,  as the time-varying system 
	\begin{equation}
	\begin{split}
	\begin{bmatrix}
	\dot{\eta}_v\\
	\dot{p}_v
	\end{bmatrix}
	&= \begin{bmatrix}
	0 & I\\
	-I & -(E_{\eta}+D_{\eta}^H)
	\end{bmatrix}\begin{bmatrix}
	\frac{\partial P_{v}}{\partial \eta_v}(\eta_v)\\
	M^{-1}_{\eta}p_v
	\end{bmatrix}+\begin{bmatrix}
	0\\
	I
	\end{bmatrix}\tau_\eta,\\
	\end{split}
	\label{eq:pH-inertial-alternative-virtual}
	\end{equation}
	This system also inherits the  passivity property of the actual one in \eqref{eq:pH-inertial-alternative}, with storage function 
	\begin{equation}
	H_{\eta_v}(\eta_v,p_v,t)=\frac{1}{2}p_v^{\top}M_{\eta}^{-1}(\eta)p_v+P_v(\eta_v),
	\end{equation}
	for any trajectory  $(\eta(t),p(t))$ and $t>t_0$ and supply rate $y_{\eta}^{\top}\tau_{\eta}$, where the output is given by $y_{\eta}=M^{-1}_{\eta}(\eta)p_v$.
	
	\subsubsection{Differential passivity based control design}
	
	\begin{proposition}\label{proposition:fullstatecontroller-inertial}
		Consider system \eqref{eq:pH-inertial-alternative-virtual} and a smooth  trajectory ${x}_d=(\eta_d,p_d)\in T^*\mathcal{Q}$. Let us introduce the following error coordinates 
		\begin{equation}
		\tilde{{x}}_v:=\begin{bmatrix}
		\tilde{{\eta}}_v\\
		{\sigma_v}
		\end{bmatrix}=\begin{bmatrix}
		{\eta}_v-{\eta}_d \hfill \\
		{p}_v-{p}_{r} 
		\end{bmatrix},
		\label{eq:changeofcoordinatesError-inertial}
		\end{equation}
		and define the auxiliary momentum reference as
		\begin{equation}
		p_{r}(\tilde{\eta}_v,t):=M(\dot{\eta}_d-\phi(\tilde{\eta}_v)),
		\label{eq:auxiliarreference-inertial}
		\end{equation}
		where  $\phi:\mathcal{Q}\rightarrow T\mathcal{Q}$ is such that $\phi(0)=0$; and $\Pi_{{\eta}_v}:\mathcal{Q}\times \mathds{R}_{\geq 0}\rightarrow \mathds{R}^{n\times n}$ is a positive definite Riemannian metric tensor  satisfying the inequality
		\begin{equation}
		\begin{split}
		\dot{\Pi}_{{\eta}_v}&(\tilde{\eta}_v,t)-\Pi_{{\eta}_v}(\tilde{\eta}_v,t)\frac{\partial \phi}{\partial \tilde{\eta}_v}(\tilde{\eta}_v)\\&-\frac{\partial \phi^{\top}}{\partial \tilde{\eta}_v}(\tilde{\eta}_v)\Pi_{{\eta}_v}(\tilde{\eta}_v,t)\leq  -2\beta_{\eta_v}(\tilde{\eta}_v,t)\Pi_{{\eta}_v}(\tilde{\eta}_v,t),
		\end{split}
		\label{eq:controllaw-positionmetricinquality-inertial}
		\end{equation}							
		with $\beta_{\eta_v}(\tilde{\eta}_v,t)>0$, uniformly. Assume also that the $i$-th row of  $\Pi_{{\eta}_v}(\tilde{\eta}_v,t)$  is a conservative vector field. Consider  also the composite control law given by
		\begin{equation}
		{\tau}(x_v,x,t):={\tau}_{ff}(x_v,x,t)+{\tau}_{fb}(x_v,x,t)+ \omega,
		\label{eq:controlaw-inertial}
		\end{equation}
		where $x=(\eta,p)$, $x_v=(\eta_v,p_v)$ and
		\begin{equation}
		\begin{split}
		{\tau}_{\eta ff}&=\dot{p}_{r}+\frac{\partial P_v}{\partial \eta_v}+\big[E_\eta(\eta,p)+D^H_\eta(\eta,p)\big] M^{-1}_\eta(\eta) p_{r},\\
		{\tau}_{\eta fb}&=-\int_{0}^{\tilde{{\eta}}_v}\Pi_{{\eta}_v}(\xi,t)d\xi-{K}_dM^{-1}_{\eta}(\eta
		)\sigma_v,
		\end{split}
		\label{eq:controlaw1-inertial}
		\end{equation}
		where ${K}_d>0$ and $\omega$ is an external input. Then,
		\begin{enumerate}
			\item   system \eqref{eq:pH-inertial-alternative-virtual} in closed-loop with \eqref{eq:controlaw-inertial}  is  differentially passive from  $\delta \omega$ to $\delta \overline{y}_{\sigma_v}=M^{-1}_\eta(\eta)\delta\sigma_v$ with respect to the differential storage function
			\begin{equation}
			W_{x}(\tilde{x}_v,\delta \tilde{x}_v)=\frac{1}{2}\delta \tilde{x}_v^{\top}\begin{bmatrix}
			\Pi_{{\eta}_v^b}& 0\\
			0 & M^{-1}_\eta(\eta)
			\end{bmatrix}\delta \tilde{x}_v.
			\label{eq:design-dLCF-sigma-inertial}
			\end{equation}			
			\item the closed-loop variational dynamics   preserves the  structure of \eqref{eq:feedbackinterconnectionsystemsvariational}, with 
			\begin{equation}
			\begin{split}
			\Pi&=\text{diag}\{\Pi_{{\eta}_v}(\tilde{\eta}_v,t),M^{-1}_\eta(\eta
			)\}\\
			\Upsilon&=\text{diag}\bigg\{\frac{\partial \phi}{\partial \tilde{\eta}_v}\Pi_{{\eta}_v}^{-1},D_\eta^H(\eta,p)+K_d \bigg\},\\
			\Xi&=\begin{bmatrix}
			0  & I\\
			-I & -S_\eta^H(\eta,p)\\		
			\end{bmatrix}; \quad \Psi=\begin{bmatrix}
			0\\I
			\end{bmatrix}.		
			\end{split}
			\label{eq:virtualvariationalsystem-inertial}
			\end{equation}
		\end{enumerate}
	\end{proposition}
	
	\subsubsection{Actual system's controller}
	Here we show that the controller that was designed  previously   can be used as a trajectoy tracking controller for the actual  Hamiltonian system \eqref{eq:pH-inertial}. This is stated as follows:
	\begin{corollary}\label{corollary:Actualtrackingcontrol-inertial}
		Consider the controller \eqref{eq:controlaw-inertial}. Then, all solutions of system \eqref{eq:pH-body}  in closed-loop with the control law  $$\tau_\eta(x,x,t)=\tau_{\eta ff}(x,x,t)+\tau_{\eta fb}(x,x,t)$$
		converge exponentially to the  trajectory $x_d(t)$  with rate 
		\begin{equation}
		\begin{split}
		\beta&=\min\{\beta_{\eta_v},\lambda_{\min}\{D_\eta+K_d\}\lambda_{\min}\{M^{-1}_\eta\}\}.
		\label{eq:trackingrate-inertial} 
		\end{split}
		\end{equation}
	\end{corollary}		
	
	\section{Example: Open-frame UUV}\label{sV}
\begin{figure}[h!]
		\centering		
		\includegraphics[width=9cm]{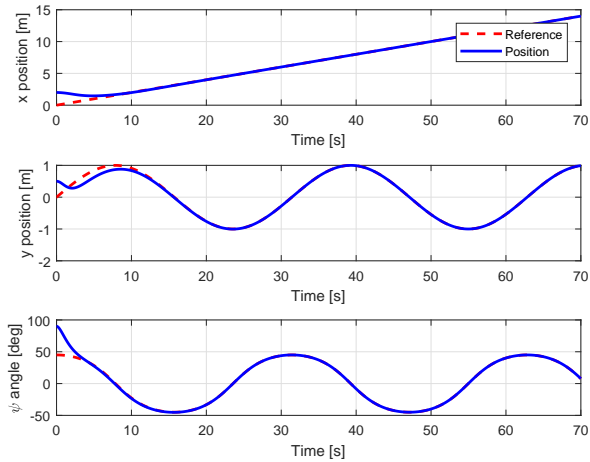}
		\caption{Vehicle's position vector $\eta$ against $\eta_d$.}
		\vspace{-0.5cm}		
		\label{fig:q}		
	\end{figure}

	We consider the example in \cite{DONAIRE20172167} which is an open frame underwater vehicle with  140 $kg$ of dry mass . The vehicle has four thrusters in an x-type configuration such that the system in fully-actuated in the degrees of freedom of interest, i.e., surge, sway and yaw. The corresponding  inertia, Coriolis and damping matrices in the body frame, respectively,  are 
	\begin{equation*}
	\begin{split}
	M&=\begin{bmatrix}
	290 & 0 & 0\\
	0 & 404 & 50 \\
	0 & 50 & 132
	\end{bmatrix},\\
	C&=\begin{bmatrix}
	0 & 0& -404v-50r\\
	0 & 0 & 290u\\
	404v+50r& -290u & 0
	\end{bmatrix},\\	
	D&=\begin{bmatrix}
	95+268|v| & 0 & 0\\0 & 613+164|u| & 0\\
	0 & 0 & 105
	\end{bmatrix}.
	\end{split}
	\end{equation*}
	Due to space limitations, we only present the performance of   controller \eqref{eq:controlaw}  in body frame, with $\phi^b(\tilde{\eta}^b_v)=\Lambda \tilde{\eta}_v^b$ where $\Lambda=\text{diag}\{0.6,0.8,0.2\}$,   $\Pi^b_{\eta_v}=\Lambda$ and $K_d=\text{diag}\{300,100,200\}$.  The time performance of system configuration $\eta$ and the desired reference $\eta_d$ is shown in Figure \ref{fig:q}. After a short transient, system's position tracks asymptotically to $\eta_d$.
	\vspace{-0.4cm} 
	
	\section{Conclusions and future research}\label{sVI}	
	In this work we have applied the v-dPBC method to solve the trajectory tracking problem in   marine craft, where  the pH structure and its workless forces have been exploited in the control design procedure. The exponential convergence to a unique predefined steady-state trajectory is guaranteed by the differential passivity property of a so-called virtual system. We have developed two families of control schemes based on body-fixed attitude and velocity measurements, one for the  pH model in   $\{b\}$ and  another for the pH model  $\{n\}$. The structure of the pH models and associated virtual systems is independent of the coordinate frame. Simulations   confirm the expected performance of the closed-loop system. 
	
	As future work,   we will use of the differential passivity property of the closed-loop system to design observers and integral actions in a differential passivity   preserving manner.
	
	
	\bibliographystyle{plain}
	\bibliography{bibliografia}

\end{document}